# Tetraquarks and pentaquarks

By Greig Cowan, University of Edinburgh and Tim Gershon, University of Warwick

Standfirst: Greig Cowan and Tim Gershon describe new types of matter called tetraquarks and pentaquarks, and discuss the outlook for understanding these particles.

# 1. Introduction

**The Standard Model and Quantum Chromodynamics**

Particle physicists are rightly proud of the Standard Model. This theory of the fundamental interactions of particles has been tested, in a variety of ways, to extraordinary precision. The most famous example regards the magnetic moment of the electron. Comparison of the most up-to-date predictions and experimental measurements of this quantity agree to within one part in a trillion, a level of precision that is unmatched elsewhere in science.

Such validations of the Standard Model are all related to one sector of the theory, however. The Standard Model is what is called a "gauge theory", in which invariance under "gauge transformations" – a kind of generalised rotation of the phases associated with quantum mechanical wavefunctions – is the cornerstone that underpins the theory. For example, in quantum electrodynamics (QED) invariance under change of phase is associated with conservation of electric charge. By requiring that this gauge symmetry also holds when the phase change can differ at each point in space and time, QED predicts the existence and properties of an associated gauge boson called the photon which mediates the electromagnetic interaction.

The best-tested part of the Standard Model is that which is related to QED, and its unification with another gauge symmetry that is responsible for the weak interaction. Since its development in the 1970s, numerous experimental observations have confirmed predictions of this electroweak sector of the theory, the most recent and spectacular example being the observation of the Higgs boson by experiments at CERN's Large Hadron Collider in 2012.

The Standard Model also explains the strong interaction through a gauge symmetry – this part of the theory is referred to as quantum chromodynamics (QCD). Somewhat enigmatically, QCD is in one respect a perfect theory, yet physicists cannot claim any precise tests of it that are comparable to those in the electroweak sector. It is perfect in the sense that, as far as is known, it is valid at all energies. At the same time, it is frustratingly difficult to use QCD to make predictions for the vast range of phenomena that are observed at low energies. This enigma is related to the large value of the QCD interaction strength – it is referred to as the strong interaction for a good reason! Consequently, interactions between particles that "feel" the strong interaction produce bound states, referred to as hadrons. These hadrons can be observed in particle detectors while their constituents are never directly observed, a property called confinement. Remarkably, there is no known way, based on first principles, to relate the underlying theory of QCD to the confined hadrons; whoever solves this problem will be able to claim one of the Millennium prizes in mathematics.

Although the methods that allow precise tests of QED cannot be applied to QCD at low energies, there are other approaches that do allow theoretical predictions to be made. One simple, but effective, method is to define empirically a confining QCD potential, as shown in Figure 1. Another, more sophisticated, technique is known as lattice QCD, and involves formulating the theory in discrete points on a four-dimensional (space and time) grid and then using powerful supercomputers to calculate quantities of interest. This approach was proposed in the 1970s, with the precision of calculations improving continually since then as more computing power becomes available. Indeed, lattice QCD has kept comparable pace to the experimental progress, so that comparison of its predictions with measurements is extremely insightful.

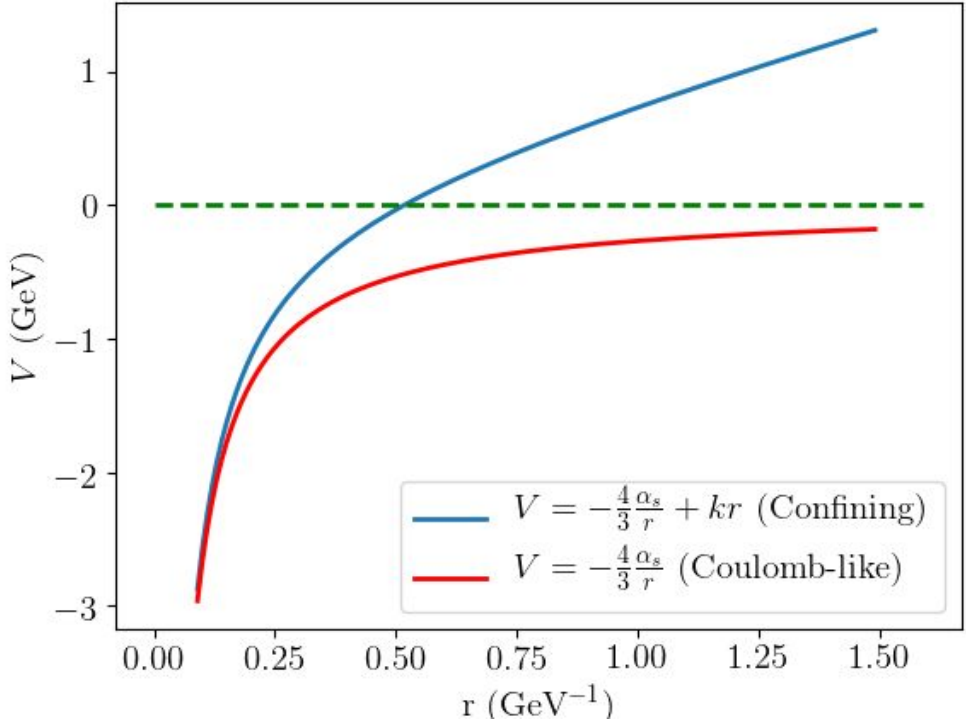

**Figure 1.** The potential (*V*) of two models as a function of the interaction radius (*r*). The Coulomb-like potential (red) asymptotically approaches zero, while the Confining potential (blue) has a linear term that allows it to cross zero, and is able to describe the long-distance behaviour of QCD. The models are generated with $\alpha_s = 0.2$, $k = 1$ GeV². Note that, following the convention in particle physics, natural units (where $\hbar = c = 1$) are used; in these units, 1 fm = $10^{-15}$ m ~ 0.2 GeV$^{-1}$.

While it may appear that lattice QCD could be a silver bullet to solve the intractability of QCD calculations, we are still some way from obtaining the computing power necessary to realise that dream. As with any other theoretical method, what you gets out depends very much on what you put in – in the language of theoretical physics, this is referred to as the choice of operators to include. Specifically, to make lattice QCD predictions of the spectrum of exotic hadrons, it is necessary to decide whether to include operators only for the quarks that are the building blocks of hadrons or also for composite objects such as diquarks (which will be explained later). To date, rather few calculations have been made including all relevant operators, but good progress is being made and continued improvements in the predictions are expected in the coming years.

**Hadrons and the quark model**

The theory of QCD states that each quark – a fundamental particle that feels the strong interaction – can come in one of three different states that are referred to as colours and labelled red, green and blue (*r, g, b*). It should be stressed that this is just a convenient labelling, and there is no connection with the colours of the electromagnetic spectrum. The QCD colour is comparable to the electromagnetic charge of QED, but instead of having any positive or negative value it is confined to only three possibilities. The colours cancel themselves out, so that a combination of *r+g+b* is called colourless. Antiquarks carry anticolour, labelled $\bar{r}$, $\bar{g}$, $\bar{b}$, and colour + anticolour, e.g. $r+\bar{r}$, is also colourless. Thus, $\bar{r}$ is equivalent to *g+b*, and so on.

The gauge bosons of QCD, known as gluons, also carry colour in one of eight possible combinations, such as $r+\bar{g}$, allowing them to interact with quarks and themselves (Figure 2). Due to the fact that the interaction is strong, quarks and gluons are continually interacting; indeed diagrams such as those in the Figure are potentially misleading, since one should always consider a haze of many interactions rather than the single vertices that are drawn. This is the fundamental reason why it is extremely challenging to make precise predictions in QCD.

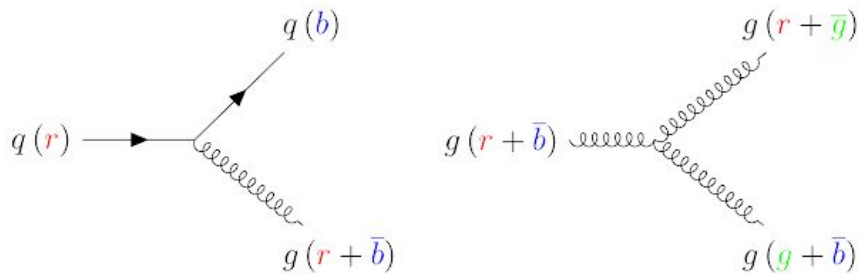

**Figure 2.** Feynman diagrams of the *qqg* and *ggg* interactions in QCD, where time travels from left to right and the interaction between particles occurs at the vertex. The labels in brackets denote the colour charges for each quark or gluon.

Nonetheless, the mere fact that only colourless combinations are observable as hadrons allows some simple predictions. It is easy to see that the simplest colourless combinations are those of a quark and an antiquark, which is known as a meson, and those of three quarks, a baryon. Given that there are six different types, or "flavours", of quark, the spectra of mesons and baryons can be predicted, at least in outline. Indeed, the experimental confirmation in 1964 of the $\Omega^-$ baryon, which had been predicted in precisely this way (at a time when only the lightest three quark flavours – up, down and strange – were known), was crucial in the development of QCD. Later, in the 1970s, the charm and beauty quarks were discovered in their quark-antiquark meson states, known as the *J/ψ* and *Υ* particles respectively, leading to the acceptance of the quark model. It was expected at the time that the 6$^{th}$ quark flavour – the top quark – would be discovered in a similar way, but it turned out to be so heavy that it decays before it can form stable hadrons. As a consequence, it was not discovered until the 1990s; also, because there are no top hadrons, we will not mention it again in the discussion that follows.

**Different types of hadrons**

Mesons and baryons are not the only possible colourless combinations of quarks and gluons. Other possibilities include so-called tetraquarks, containing two quarks and two antiquarks, and pentaquarks, containing four quarks and one antiquark (or the antimatter equivalent: four antiquarks and one quark[1]). Two or three gluons can potentially also form a colourless object, referred to as a glueball. Moreover, there could also be hybrid states, which are some mixture of conventional hadron and glueball. According to the theory of QCD, one would expect that it is possible to produce these "exotic" types of hadrons in particle physics experiments. Until 2003, however, there was no clear evidence for exotic hadrons, but since then there has been an explosion in the rate of observations of new hadronic states whose properties indicate that they are combinations of at least four or five quarks and antiquarks. A recent high-profile example, illustrated in Figure 3, was the discovery of two pentaquark candidates in data collected by the LHCb experiment at the CERN Large Hadron Collider (LHC), using the world's largest sample of *b*-quark hadron decays.

---

[1] Since the physics being discussed is identical for matter and antimatter, in the remainder of this work only matter versions of particles will be mentioned but it should be understood that antimatter equivalents are possible and that both are often studied simultaneously in experiments.

In the following chapters we explain the background to this new era of exotic hadron spectroscopy, discussing how the newly observed states are produced and identified in experiments. Specific examples of recent discoveries are given to highlight the current directions in research into tetraquarks and pentaquarks. We also summarise emerging theoretical approaches that may explain the observed hadron spectrum and the dynamics of multiquark states, and give an outlook for future measurements that should help to clarify which, if any, is the correct underlying description.

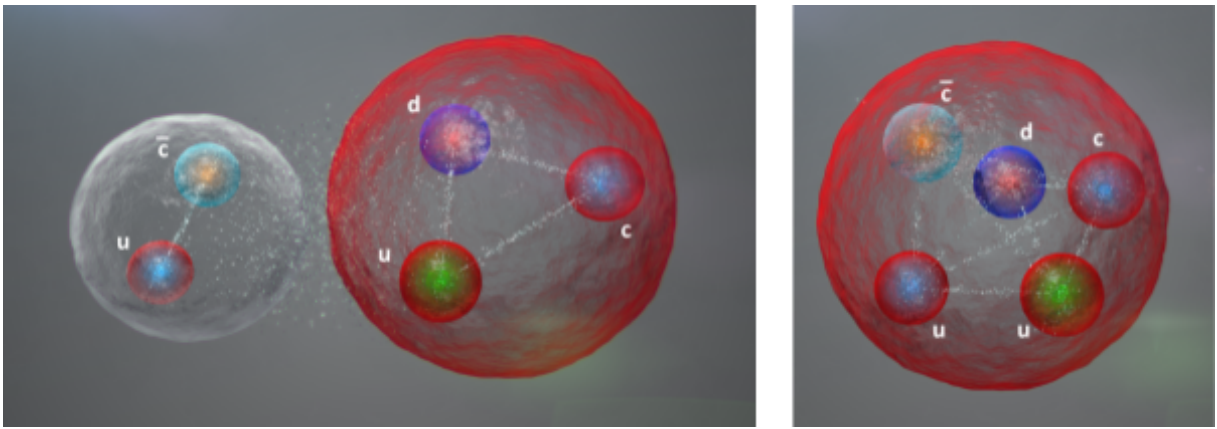

**Figure 3.** Possible quark combinations to make mesons, baryons and pentaquarks. Copyright CERN.

## 2. Background

**Understanding hadrons through their quantum numbers**

Given that we never observe the constituents of hadrons, how can we know whether a particle observed in experiment is a conventional meson or baryon, rather than an exotic hadron? A key piece of information that is used to classify hadrons is their quantum numbers. These are a set of fundamental properties that distinguish different types of particles. Good examples are the total angular momentum (denoted $J$ and within the field of spectroscopy usually, though a bit confusingly, called "spin") and parity ($P$) quantum numbers, where parity is a label (either + or −) that indicates how a particle's wavefunction behaves when all spatial coordinates are inverted.

Consider a meson, *i.e.* a quark-antiquark system. Quarks are spin-1/2 particles (fermions) and therefore if there is no orbital angular momentum ($L$) between them the total angular momentum is equal to the intrinsic spin ($S$), which is either zero (if the spins are anti-aligned) or 1 (aligned). It is a fundamental property of fermions that particle and antiparticle have opposite parity, and therefore $P=-$. Thus, the possible spin and parity quantum numbers are $J^P = 0^-$ and $1^-$, referred to as pseudoscalar and vector respectively. If there is one unit of orbital angular momentum between quark and antiquark, the parity is flipped and the total spin has three possible values following the rule for adding angular momentum in quantum mechanics $J = L \oplus S$. This results in four states with $J^P = 0^+, 1^+, 1^+, 2^+$. States with higher spin have higher energy, and hence higher mass, so a distinctive pattern of states is predicted. If particles are observed that do not fit into this pattern, that could be a signature of their exotic nature.

Another relevant quantum number is known as charge conjugation (*C*). Similar to parity, *C* can be either + or −, depending on how the wavefunction behaves when all particles are replaced with antiparticles. For the *C* quantum number to be well-defined, a particle must be an eigenstate of charge conjugation, which means that it must be its own antiparticle. Among hadrons, *C* is therefore relevant only for mesons where quark and antiquark have the same flavour, often referred to as "onia" in the case of heavy-quarks (*c* and *b*). The rule relating *C* to the total spin is $C=(-)^{L+S}$ (here a simple addition of *L* and *S* is used). Consequently for onia the $L=0$ states have $J^{PC} = 0^{-+}$ and $1^{--}$, while the orbitally-excited states are $0^{++}, 1^{++}, 1^{+-}$ and $2^{++}$. For the charmonia ($c\bar{c}$) states, these six combinations correspond to particles known as $\eta_c, J/\psi, \chi_{c0}, \chi_{c1}, h_c$ and $\chi_{c2}$ respectively. (The properties of these, and other hadrons referred to throughout the text, are summarised in Table 1 for convenience.)

Another type of labelling of states, familiar from atomic spectroscopy, is sometimes used: $n^{2S+1}L_J$, where the notation S, P, D, F is used to correspond to orbital angular momentum values $L = 0, 1, 2, 3$, respectively. In this convention the same six charmonia states are $^1S_0, ^3S_1, ^3P_0, ^3P_1, ^1P_1, ^3P_2$, where all have the principal quantum number $n = 1$. So called radial excitiations, where $n = 2$ or higher, are also of interest and are sometimes referred to with one of two further notations, *e.g.* $\psi(2S)$ or $\psi(3686)$, where the values in parentheses are the $nL$ quantum numbers and the mass in units of MeV, respectively. This notation is often used for open flavour hadrons such as mesons involving quark and antiquark of different flavours.

Determination of the quantum numbers of hadrons is essential to understand their quark content and can be crucial to establish their nature as either exotic or conventional states. For example, certain combinations of *J, P* and *C* cannot be formed by quark-antiquark mesons and are therefore explicitly exotic: these include $0^{--}, 0^{+-}$ and $1^{-+}$. The quantum numbers can be measured by studying how a particle is produced, or (more often) what it decays to, since all of *J, P* and *C* are conserved in both strong and electromagnetic interactions. Therefore, if the *P* and *C* values for the final state of a strong or electromagnetic decay are known, then the properties of the decaying hadron can be elucidated.

The weak interaction violates the conservation of both *P* and *C* quantum numbers, but weak decays are easily identified since the much smaller interaction strength means that these only occur when no other decay is possible, leading to measurably long lifetimes. This is the case for the ground state of each open flavour hadron, since the weak interaction is also the only force in the Standard Model that allows changes of quark flavour. The open flavour charm and beauty hadrons in Table 1 have lifetimes of the order of 1 ps, *i.e.* $10^{-12}$ s; strange hadrons have lifetimes up to around $10^{-8}$ s.

Conservation of quark flavour can also be used to identify exotic hadrons. For example, if a meson contains a charm quark-antiquark pair, it should be neutral. Thus if a hadron decays strongly to a final state containing a $J/\psi$ meson plus another charged particle, that hadron cannot be a conventional meson. Another quantum number known as isospin can similarly provide important information. Isospin is related to the fact that the masses of the up and down quarks are nearly identical, and therefore they are interchangeable as regards their interactions with the strong force. Hadrons containing $u\bar{u}$ or $d\bar{d}$ components therefore tend to mix together to form states with isospin 1 (such as pions and ρ mesons) or zero (η and ω mesons). Isospin is approximately conserved in strong decays, so the decay rate for a conventional $c\bar{c}$ meson to a final state with non-zero isospin should be highly suppressed.

**Table 1.** Conventional mesons and baryons, referred to in the text, with quantum numbers, indicative mass values and quark content. The charge conjugation quantum number is relevant only for mesons composed of quark and antiquark of the same flavour. The quark content given for the η and ω mesons neglects mixing with $s\bar{s}$ states.

| Particle ($J^{P(C)}$) | Typical mass (GeV) | Quark content |
|---|---|---|
| $\pi^+$, $\pi^0$, η ($0^{-(+)}$) | 0.14 (π), 0.55 (η) | $u\bar{d}$, $u\bar{u} - d\bar{d}$, $u\bar{u} + d\bar{d}$ |
| $\rho^+$, $\rho^0$, ω ($1^{-(-)}$) | 0.78 | $u\bar{d}$, $u\bar{u} - d\bar{d}$, $u\bar{u} + d\bar{d}$ |
| p, n (½$^+$) | 0.94 | $uud$, $udd$ |
| $K^+$, $K^0$ ($0^-$) | 0.50 | $u\bar{s}$, $d\bar{s}$ |
| $K^{*+}$, $K^{*0}$ ($1^-$) | 0.89 | $u\bar{s}$, $d\bar{s}$ |
| Λ (½$^+$) | 1.1 | $uds$, $usc$, $dsc$ |
| $D^+$, $D^0$, $D_s^+$ ($0^-$) | 1.9 | $c\bar{d}$, $c\bar{u}$, $c\bar{s}$ |
| $D^{*+}$, $D^{*0}$, $D_s^{*+}$ ($1^-$) | 2.0 | $c\bar{d}$, $c\bar{u}$, $c\bar{s}$ |
| $\Lambda_c^+$, $\Xi_c^+$, $\Xi_c^0$ (½$^+$) | 2.3 - 2.5 | $udc$, $usc$, $dsc$ |
| $\eta_c$, $J/\psi$, $\chi_{c0}$, $\chi_{c1}$, $h_c$, $\chi_{c2}$ ($0^{-+}, 1^{--}, 0^{++}, 1^{++}, 1^{+-}, 2^{++}$) | 2.9 - 3.6 | $c\bar{c}$ |
| $B^+$, $B^0$, $B_s^0$ ($0^-$) | 5.3 | $u\bar{b}$, $d\bar{b}$, $s\bar{b}$ |
| $\Lambda_b^0$, $\Xi_b^0$, $\Xi_b^-$ (½$^+$) | 5.6 - 5.8 | $udb$, $usb$, $dsb$ |
| $\eta_c$, Y, $\chi_{b0}$, $\chi_{b1}$, $h_b$, $\chi_{b2}$ ($0^{-+}, 1^{--}, 0^{++}, 1^{++}, 1^{+-}, 2^{++}$) | 9.4 - 9.9 | $b\bar{b}$ |

**Discovery of the *J/ψ* meson**

Among the different charmonia states introduced above, the *J/ψ* is particularly important for studies of tetraquarks and pentaquarks. The discovery of this state provides useful insight into why this should be so. As already mentioned, the *J/ψ* was the first particle containing a charm quark to be discovered. Since it is heavier than the $\eta_c$ charmonium meson and the open charm $D^+$ and $D^0$ states, why were they not discovered first?

The *J/ψ* was discovered contemporaneously at two experiments: one at the Brookhaven National Laboratory (BNL) using a proton beam impinging on a beryllium target, and the other at the Stanford Linear Accelerator Center (SLAC) in which electron and positron beams collided. (This dual discovery is the reason behind the unusual double name, which is unique for this particle.) In $e^+e^-$ collisions, the incoming particles annihilate with each other to produce a virtual photon, and therefore only particles with the same quantum numbers as the photon ($J^{PC} = 1^{--}$) can be produced. Due to conservation of flavour quantum numbers in electromagnetic interactions, open charm states can only be produced in pairs, and therefore only at centre of mass energies higher than twice the $D$ mass, somewhat above the $J/\psi$ threshold (see Table 1). Moreover, strong interaction decays of the $J/\psi$ to lighter hadrons occur only through interactions that are suppressed by the so-called Okubo-Zweig-Iizuka (OZI) rule, named after the three physicists who discovered it. This occurs since the $c\bar{c}$ pair must annihilate into a minimum of three gluons: one is not possible since the onia states are colourless; two gluons could be colourless but would have $C=+$, while the $J/\psi$ has $C=-$. Thus, among the possible decays of the $J/\psi$ meson, those to $e^+e^-$ or $\mu^+\mu^-$ final states occur quite frequently (in particle physics terminology, they have relatively large branching fractions), though decays to hadrons are also possible. In any case, all of these final states can be detected in experiments at $e^+e^-$ colliders, since there are no other particles produced in the collision, making the events relatively simple to analyse.

The situation for hadron collisions, including those in the BNL experiment, is quite different. Particles with any quantum numbers can be produced through quark-antiquark or gluon-gluon interactions. However, the collisions tend to produce large numbers of additional particles, making it difficult to observe signals from decays of new states since there are so many possible random combinations of particles (known as background). In order to suppress this background it is necessary to look for decays involving types of particles, such as leptons, that are produced less often when the signal is absent and that have distinctive signatures in the detector. The leptonic $e^+e^-$ or $\mu^+\mu^-$ final states are therefore ideal discovery channels for new particles, but among the open charm and charmonia states, only the $J/\psi$ has a large branching fraction to them. Thus, the BNL experiment could observe the $J/\psi \rightarrow e^+e^-$ decay, but not see any other new particles.

With this in mind, the answer to the question of why the $J/\psi$ was the first particle containing charm to be discovered becomes clear. In the SLAC experiment, only the $J/\psi$ was produced although any charm state could in principle have been detected. In the BNL experiment all charm and charmonia states were produced, but only the $J/\psi$ could be detected. It is this distinctive signature of the $J/\psi$ which has made it central to the study of heavy flavoured hadrons.

**Resonances, Breit-Wigner lineshape and phase changes**

Another reason why the $J/\psi$ signal is so distinctive is related to the OZI-suppression mentioned above. As the number of different final states available for decay is reduced, and the rates to many of the possible decay channels suppressed, the $J/\psi$ has a longer lifetime than could otherwise be expected. The concept of "lifetime" for such a particle is, however, not necessarily appropriate since it still decays essentially instantaneously. Instead, the $J/\psi$ is said to have a narrow width, meaning that the spread in the possible mass values is small. The relation between width and lifetime occurs due to Heisenberg's uncertainty principle, $\Delta E \Delta t \geq \hbar/2$ where $\Delta E$ and $\Delta t$ are uncertainties in energy (or, equivalently, mass) and time, and $\hbar$ is the reduced Planck's constant. Thus, particles with very short lifetimes, typically of order $10^{-23}$ s or less, appear as peaks in mass or energy distributions with a finite width. Such states are referred to as resonances.

In a particle or nuclear physics experiment that is studying a particular reaction, resonances show up as enhancements of the reaction cross-section as a function of the collision centre-of-mass energy or as a "bump" in the invariant mass spectrum of the decay products of the reaction. In the simplest case, the resonant amplitude is described by a so-called Breit-Wigner function, which has the form

$$A(E) \propto \frac{1}{m^2 - E^2 + im\Gamma(E)},$$

where $E$ is the energy or invariant mass of the process, $m$ is the resonance mass and $\Gamma$ is the resonance width. As is clear from the above expression, the Breit-Wigner function is complex-valued, having both a magnitude and a phase. The rate (production cross-section or decay probability) for a particular process is proportional to the square of the magnitude, $|A(E)|^2$. Figure 4 shows a graphical representation of the Breit-Wigner function.

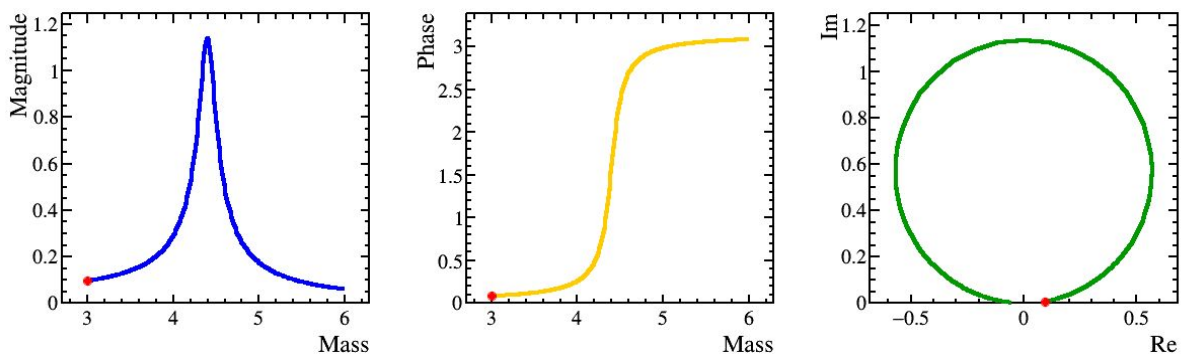

**Figure 4.** Graphical representation of how a Breit-Wigner function for a particular resonance, with $m = 4.45$ GeV and $\Gamma = 0.20$ GeV, varies with the the invariant mass of its decay products. The blue (left) curve shows how the magnitude of the complex amplitude rises rapidly towards the mass of the resonance before dropping down again, giving a characteristic "resonance peak". Simultaneously, the yellow (middle) curve shows the resonance phase going through a rapid 180° variation as we scan across the mass. Since the Breit-Wigner function is complex valued, it is insightful to recast the magnitude and phase as real and imaginary parts and to plot them on the Argand plane, as shown by the green (right) curve. This demonstrates a circular pattern, moving anti-clockwise as the mass increases. This behaviour provides a characteristic fingerprint that can be used to determine if experimental data support the hypothesis that a particular bump is truly a resonance. Animation available online at http://iopscience.iop.org/book/978-0-7503-1593-7

**Controversy over claimed light pentaquarks**

The search for exotic hadrons such as tetraquarks and pentaquarks has a somewhat chequered history, which is important to keep in mind when considering more recent developments. In particular, during the early 2000's several fixed-target scattering experiments observed enhancements above the expected background in the mass spectra of the processes they were studying (*e.g.,* positive kaons scattering from a proton target). Many of these enhancements appeared at approximately 1.54 GeV with narrow width, and were interpreted as evidence for an exotic pentaquark state, denoted $\Theta^+$, with quark content $uudd\bar{s}$ that had earlier been predicted by theoretical models.

These results generated huge theoretical interest and initiated new experimental searches for related light-quark exotic states. However, updated experimental results involving larger data samples and more detailed investigation did not confirm the earlier claims. While some of the claimed new states that appeared during this era have not been completely disproven, it is now generally accepted that the majority of the ostensible signals were fake, caused by a combination of statistical fluctuations and unaccounted-for systematic uncertainties. This stands as a clear reminder about the need for extreme care in the analysis of experimental data when claiming observations of new resonances.

While the excitement over the $\Theta^+$ has receded, there remain some interesting observations of apparently non-$q\bar{q}$ mesons with masses below 2 GeV. For example, in the scalar meson ($J^{PC} = 0^{++}$) sector, more states have been observed than are expected from simple $q\bar{q}$ mesons. It is possible that one or more of these states may be due to four-quark combinations or glueballs. Similarly, there are several candidate hybrid mesons that have manifestly exotic quantum numbers, such as the $\pi_1(1400)$ and $\pi_1(1600)$ with $J^{PC} = 1^{-+}$. However, the interpretation of experimental data on light-quark states is extremely complicated owing to the large widths of these resonances and the fact that all states with the same quantum numbers mix quantum mechanically. Thus, it has not been possible to isolate which, if any, of these resonances have an exotic component.

**Heavy flavour hadrons, and the experimental facilities used to study them**

Studies of hadrons containing charm and beauty quarks (so-called heavy flavours) have a two-fold advantage for understanding QCD compared to the light-quark hadrons. First, it is simpler and more reliable to perform theoretical calculations to predict the spectrum of expected states using well-understood tools such as non-relativistic quantum mechanics and lattice QCD. This allows theory to guide where experiments should search for resonances. Secondly, heavy-quark hadrons often decay to final states with leptons, which are ideal for experimental detection. Together, these aspects have led to a very detailed understanding of the spectrum of charmonium mesons below the threshold for production of pairs of *D* mesons at approximately 3.73 GeV. This is clear from the lower section of Figure 5 where there is excellent agreement between the properties of theoretically predicted states and those that have been experimentally well-established.

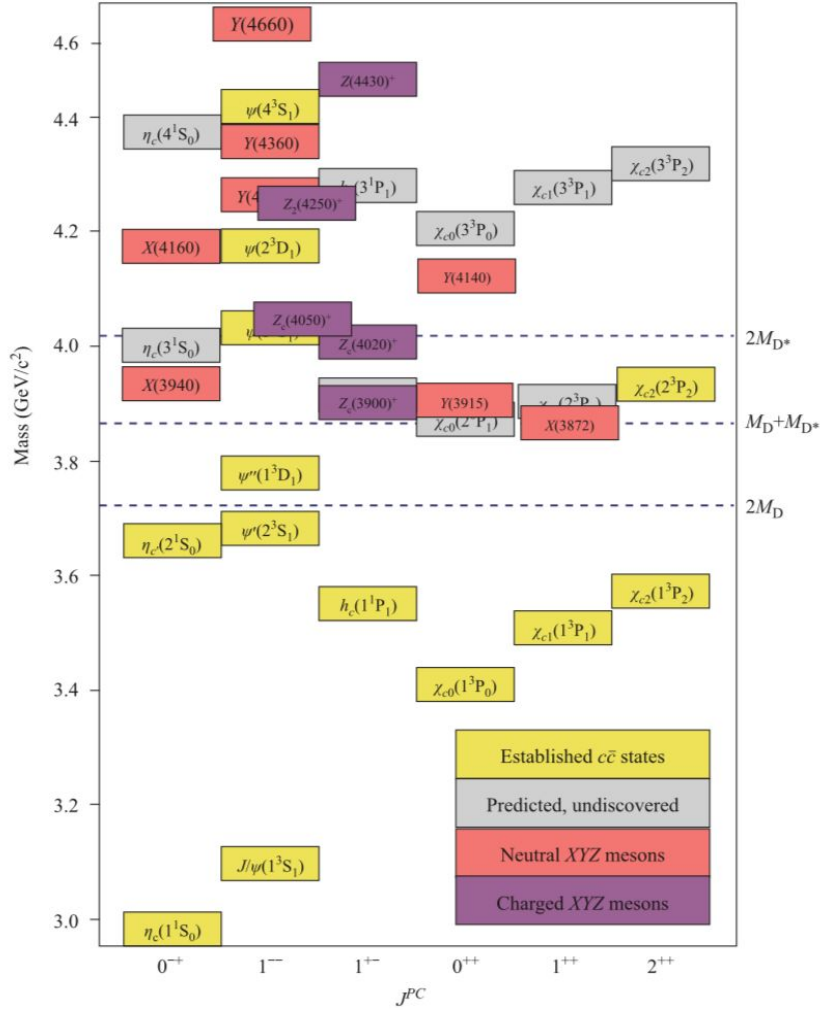

**Figure 5.** Spectrum of charmonium mesons, including experimentally well-established states, those that have been theoretically predicted but undiscovered and the new neutral and charged XYZ tetraquark-like candidates. Reproduced from *Front. Phys.* **10**, 101401 (2015).

Since *b* and *c* hadrons are heavy, they are difficult to produce in particle collisions. Therefore, specialist facilities operating at high energies are necessary to make them in sufficiently large numbers that physicists can measure their properties precisely. Table 2 summarises some of the key past, current and future experimental facilities for studies of exotic heavy flavoured hadrons from around the world. These facilities use various different beams of particles to produce $b\bar{b}$ and $c\bar{c}$ quark pairs. Specialised detectors are constructed around the interaction points to measure the directions and energies of the final state particles that emerge from the collisions, providing an electronic snap-shot that allows physicists to kinematically reconstruct the intermediate processes. For example, at the CERN Large Hadron Collider (LHC) a significant fraction of proton-proton collisions produce pairs of $b\bar{b}$ quark pairs, which subsequently combine with other light-quarks via a process called hadronisation to form *b*-flavoured mesons and baryons. If the produced hadrons are not in a ground state, they decay instantaneously through strong and electromagnetic interactions until no further such transitions are possible. The ground-state particles then travel of the order of 1 cm in the LHC detectors before decaying, through the weak interaction, to lighter hadrons (pions, kaons, protons) and/or leptons. It is these final state particles that ultimately deposit energy in the detector subsystems, which is used as the starting point to reconstruct the chain of decays.

**Table 2.** Some of the particle physics experiments that have contributed significantly to knowledge of the exotic hadron spectrum. Future experiments that are expected to have a major impact are also included. The Tevatron and LHC accelerators provide symmetric beam collisions, as do CESR and BEPC. At BaBar, Belle and Belle II asymmetric electron and positron beam energies are used. At PANDA an antiproton beam impinges on one of several possible fixed targets. A photon beam striking fixed targets is used for experiments at JLab.

| Experiments | Laboratory | Accelerator facility | Production process (centre-of-mass energy) | Operational period |
|---|---|---|---|---|
| CDF/D0 | Fermilab, USA | Tevatron | $p\bar{p} \to b\bar{b}\,X$ (2 TeV) | 1987-2011 |
| BaBar | SLAC, USA | PEP-II | $e^+e^- \to Y(4S) \to B\bar{B}$ (10.6 GeV) | 1999-2008 |
| Belle | KEK, Japan | KEKB | $e^+e^- \to Y(4S) \to B\bar{B}$ (10.6 GeV) | 1999-2010 |
| CLEO-c | Cornell, USA | CESR | $e^+e^- \to c\bar{c}$ (3.7-4.2 GeV) | 2003-2008 |
| BESIII | IHEP, China | BEPC | $e^+e^- \to c\bar{c}$ (3.0-4.6 GeV) | 2008-ongoing |
| ATLAS/CMS/ LHCb | CERN, Switzerland | LHC | $pp \to b\bar{b}\,X$ (7-13 TeV) | 2010-ongoing |
| Belle II | KEK, Japan | Super-KEKB | $e^+e^- \to Y(4S) \to B\bar{B}$ (10.6 GeV) | 2018-2025 |
| GlueX/CLAS12/ E12-16-007 | JLab, USA | CEBAF | $\gamma p \to c\bar{c}\,X$ (4-5 GeV) | 2017-ongoing |
| PANDA | GSI, Germany | FAIR | $p\bar{p} \to c\bar{c}\,X$ (2.9-5.5 GeV) | 2025- |

Over the past 20 years the capability of accelerators and detectors has improved dramatically, so that much larger samples of *b*-hadrons have become available and the approach of studying charmonium states produced via such decays has become increasingly more powerful. This method tends to have lower background rates compared to cases where the charmonia are produced directly in collisions. Moreover, once the *b*-hadron momentum has been determined, the initial state is strongly constrained giving excellent potential to determine the quantum numbers of any exotic state. As an example, Figure 6 shows the Feynman diagram for the process $B^+ \to \psi(2S)K^+$, where the $\bar{b}$ quark undergoes a weak decay to a $\bar{c}$ quark. The intermediate $W$ boson then produces $c$ and $\bar{s}$ quarks, giving rise to the $\psi(2S)$ meson and a charged kaon in the final state. The $\psi(2S)$ meson subsequently decays, with the final state of two oppositely charged pions and a $J/\psi$ meson (which itself decays to two muons or two electrons) being particularly relevant for the following discussion. As mentioned earlier, the experimental detection of leptons that are displaced from the *pp* interaction gives a very effective signature to isolate *b*-hadron decays from the large background of less-interesting particles that arise in each collision.

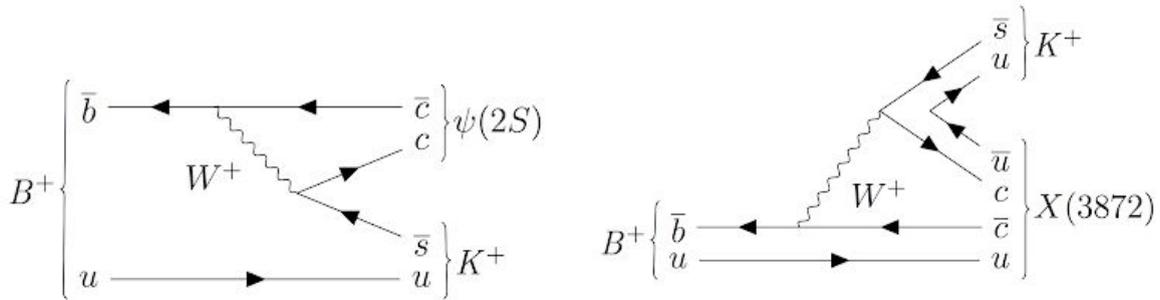

**Figure 6.** Feynman diagrams showing two different decays of a $B^+$ meson. On the left is the decay to a final state containing the conventional *ψ(2S)* charmonium state and on the right is the decay to the exotic *X(3872)*. The *ψ(2S)* and *X(3872)* both decay to $J/\psi\pi^+\pi^-$, such that the same particles are in the final state of both $B^+$ meson decays.

**Discovery of the *X(3872)***

The result that kicked off the current era of exotic hadron studies was the observation in 2003 of a particle called the *X(3872)*, where "X" in the name signifies uncertainty about the nature of the state. This discovery was made using data recorded by the Belle experiment when studying the $J/\psi\pi^+\pi^-$ invariant mass spectrum of $B^+$ mesons decaying to $J/\psi\pi^+\pi^-K^+$. The original Belle experiment data, displayed in Figure 7, show the very clear signature of the well-known ψ(2S) meson as well as an unexpected second narrow peak. This indicates another resonance that decays to the same final state, $J/\psi\pi^+\pi^-$.

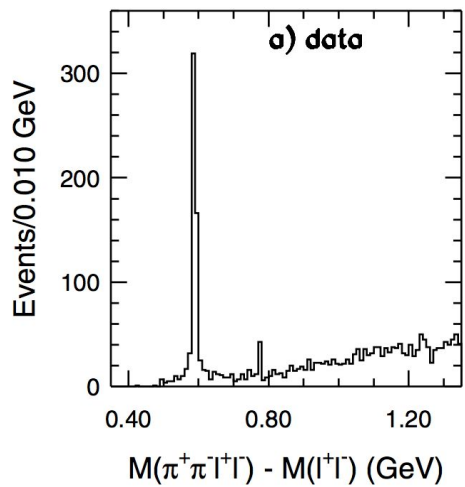

**Figure 7.** Distribution of the difference between $\pi^+\pi^-l^+l^-$ and $l^+l^-$ invariant masses of $B^+ \to J/\psi\pi^+\pi^-K^+$ candidates, where $J/\psi \to l^+l^-$, *l* being *μ* or *e*. The narrow peaks near 0.6 and 0.8 GeV correspond to the ψ(2S) and *X(3872)* states, respectively. Reproduced from *Phys. Rev. Lett.* **91** (2003) 262001.

A feature of many exotic hadron observations is that they are seen by only one experiment and/or in only one decay mode. However, the *X(3872)* is refreshingly different in this regard:

shortly after its discovery by Belle it was confirmed by the BaBar experiment, and also observed to be produced in $p\bar{p}$ collisions by the CDF and D0 experiments. Other decay modes, including $J/\psi\gamma$ and $D^0\overline{D}^{*0}$ were observed. Soon after the LHC came on line, all of the ATLAS, CMS and LHCb experiments were able to measure its production rate in $pp$ collisions. This wealth of new and somewhat mysterious data led to huge interest within the particle physics community. The latest results include precise measurement of the *X(3872)* mass (3871.69 ± 0.17 MeV) and unambiguous determination of its quantum numbers ($J^{PC} = 1^{++}$).

Even though the *X(3872)* is the most well-studied exotic hadron its true nature remains a mystery. One may wonder, given its quantum numbers and the fact that its mass is in the same region as other conventional $c\bar{c}$ mesons, could it not simply be the previously unobserved charmonium state $\chi_{c1}(2P)$? This explanation does not appear to fit the data in several ways. For example, detailed analysis of the discovery decay mode $X(3872) \to J/\psi\pi^+\pi^-$ shows that the $\pi^+\pi^-$ pair comes predominantly from the $\rho^0$ meson, indicating violation of isospin symmetry that is not expected for a conventional charmonium meson. It would also be hard to explain why $\chi_{c1}(2P)$ width is so small, as experiments have set an upper limit on the *X(3872)* width of < 1.2 MeV, comparable to the value for the $\chi_{c1}(1P)$ state even though decays to open charm are allowed.

If the *X(3872)* is not a pure $c\bar{c}$ state, the simplest explanation would appear to be that it is composed of two quark and two antiquarks ($c\bar{c}u\bar{u}$) and produced in $B^+$ decay as shown in Figure 6 (right). The question then becomes precisely how its constituent parts are bound together. One popular hypothesis is that it is a molecule of $D^0$ and $\overline{D}^{*0}$ mesons bound together via pion exchange, just as the deuteron is considered to be a bound state of a proton and neutron. This model explains the observed isospin violation, since the *X(3872)* mass is right at the threshold to make a $D^0$-$\overline{D}^{*0}$ pair, which is slightly but significantly below the corresponding threshold involving charged *D* mesons. A molecular state may be expected to have production rates smaller than those observed in $pp$ and $p\bar{p}$ collisions, but this could be explained if the *X(3872)* wavefunction includes an admixture of $\chi_{c1}(2P)$. An alternative explanation is that the *X(3872)* may be a heavy tetraquark, with quarks and antiquarks tightly bound into a colour singlet object through so-called diquarks: $[cu][\overline{cu}]$. This model allows the decay to $J/\psi\rho^0$, but predicts the existence of charged partners that should decay to $J/\psi\rho^+$, which have not yet been found.

The concept of a diquark deserves further explanation. In diquark models, pairs of quarks are considered to bind together to form a single object that interacts with the third quark (in a baryon) or an anti-diquark (in a tetraquark) through gluon exchange. In QCD, the sum of two colours is equivalent to anticolour (*e.g. g+b* corresponds to $\bar{r}$) so in this respect diquarks behave like antiquarks. Thus, in diquark models, the QCD modelling of either a baryon or a tetraquark can be handled in the same way as for mesons.

Existing measurements are not able to distinguish clearly between these molecular and tightly bound diquark models, but continued experimental studies of the *X(3872)* properties will help to understand its structure. Another approach is to look for other exotic hadrons and see if a pattern emerges. Indeed, since 2003 many new states have been observed in the charmonium and bottomonium systems that cannot be matched to predictions from the quark model. These thirty-or-so hadrons have been found above the corresponding threshold for production of pairs of *D* or *B* mesons, as can be seen (for the $c\bar{c}$ case) by the various coloured boxes above 3.73 GeV in Figure 5. In the following sections we will explain more about how some of these exotic states were discovered and their properties measured, and describe the ongoing research that aims to provide understanding of their true nature.

## 3. Current directions

**Tetraquarks**

While there is strong evidence that the *X(3872)* has an exotic nature, it is hard to completely discount the possibility that it is related to conventional charmonia. In order to be certain of the existence of exotic hadrons, a "smoking gun" signature is required. As mentioned earlier, this can be provided by a charged charmonium-like resonance, the first of which to be discovered was the $Z(4430)^-$ state, initially seen by Belle as a peak in the invariant mass of $\psi(2S)\pi^-$ combinations found in the decays of *B* mesons: $B^0 \to \psi(2S)K^+\pi^-$ with $\psi(2S) \to \mu^+\mu^-$. The first claims of discovery were, however, not universally accepted – the BaBar experiment analysed a similarly-sized dataset using a complementary method and found no evidence for contributions from exotic hadrons. A major concern was that the Belle analysis had been too simplistic, without accounting properly for so-called reflections from excited kaon ($K^*$) resonances decaying to $K^+\pi^-$. Since various $K^{*0}$ states with different spins can contribute, a complicated interference pattern that could fake a peak in $\psi(2S)\pi^-$ invariant mass may arise.

Although Belle reanalysed their data with more sophisticated approaches, this controversy was not finally resolved until 2014 when LHCb released results based on a sample of *B* meson decays an order of magnitude larger than those available to Belle and BaBar. Rather than simply inspecting the $\psi(2S)\pi^-$ invariant mass, the LHCb physicists developed an amplitude model that made use of the full kinematic information available in the $B^0 \to \psi(2S)K^+\pi^-$ decay to increase the sensitivity of their data analysis to exotic hadron contributions. As an example, in a three-body decay involving only spin-0 particles it can be shown via energy-momentum conservation that only two variables are needed to describe the decaying system fully. These variables are typically taken to be the squared invariant masses of the two-body combinations of the final state particles. Making a scatter plot of the data in this two-dimensional plane leads to a characteristic diagram known as a Dalitz plot, named after physicist Richard Dalitz. In the $B^0 \to \psi(2S)K^+\pi^-$ process, these variables can be taken to be $m(K^+\pi^-)^2$ and $m(\psi(2S)\pi^-)^2$; however, since the decay involves a spin-1 ($\psi(2S)$) particle, an additional two variables describing the orientation of the $\psi(2S) \to \mu^+\mu^-$ decay are required to encapsulate the full kinematics. By exploiting all the kinematic information available in the decay, overlapping states can be resolved so long as they have different quantum numbers.

The crucial point about the amplitude model is that it is constructed to describe all of the $K^{*0}$ components that contribute to the $B^0 \to \psi(2S)K^+\pi^-$ process, in addition to any exotic states. The conventional $K^*$ resonances have been well studied in previous experiments so their masses, width and $J^P$ quantum numbers are known. After accounting for effects due to background contamination and reduced efficiency for detecting the final state particles in different regions of the 4D space, the LHCb physicists found that they were unable to describe their data using only $K^{*0}$ resonances in the amplitude model. Only through the introduction of an exotic $Z^- \to \psi(2S)\pi^-$ resonance was the amplitude fit able to give a good description of the data. Thus, the existence of the $Z(4430)^-$ state was finally confirmed, with overwhelming significance.[2] It is noteworthy, however, that the value of the $Z(4430)^-$ width obtained from the amplitude model is much larger than that obtained by the original Belle analysis. This underlines the importance of very careful analysis to obtain reliable results.

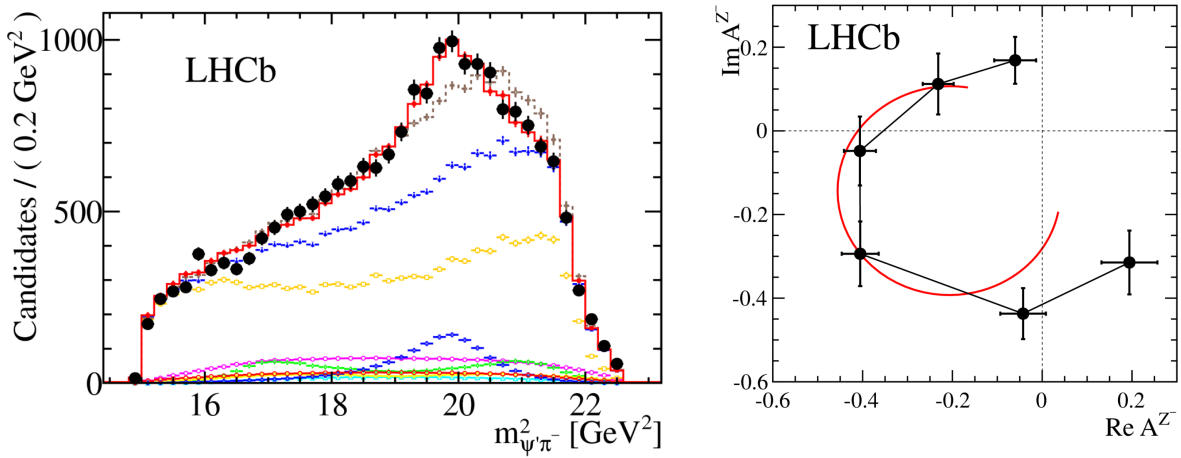

**Figure 8.** (left) Squared invariant mass spectrum of $\psi(2S)\pi^-$ combinations in $B^0 \to \psi(2S)K^+\pi^-$ decays. The blue enhancement near 20 GeV² shows the contribution from the $Z(4430)^-$ tetraquark candidate to the total decay rate, given by the solid red line and agreeing well with the data points shown in black. (right) The data points show the variation of the $Z(4430)^-$ decay rate in the Argand plane, displaying the characteristic circular behaviour of a resonance.

The results of the LHCb amplitude fit can be seen in Figure 8, where the projection of the data and final model onto the $m(\psi(2S)\pi^-)^2$ dimension is shown. Here, the $Z(4430)^- \to \psi(2S)\pi^-$ component is visible as the Breit-Wigner-like peak around 20 GeV², which is interfering with the substantially larger contributions from $K^{*0}$ resonances, shown by the broad coloured histograms. The total fit result is given by the solid red line. The dotted brown line shows the result of the fit without any exotic $\psi(2S)\pi^-$ components, which clearly disagrees with the data in the region around 20 GeV². The mass, width and quantum numbers of the $Z(4430)^-$ were measured to be $4475 \pm 7 \pm 20$ MeV, $172 \pm 13 \pm 35$ MeV and $J^P = 1^+$ respectively, where the first uncertainty is statistical and the second due to systematic uncertainties in the measurement.³

---

[2] Particle physicists measure significance in terms of numbers of standard deviations (usually denoted $\sigma$). Significance greater than $5\sigma$ is usually required to claim observation. The significance of the $Z(4430)$ in the LHCb analysis was found to be over $13\sigma$, corresponding to a vanishingly small probability (1 in $10^{38}$) that the result is a fake caused by random fluctuations.

These results clearly establish the need for contributions beyond those from conventional $K^{*0}$ resonances in the $B^0 \to \psi(2S)K^+\pi^-$ amplitude model. However, there remains a possibility that the $Z(4430)^-$ signal is not due to an exotic resonance, but rather could be caused by a dynamical effect. For example, it is possible that a $B$ meson decays to a charm meson-anticharm meson pair and a kaon, and that a strong interaction rescattering process (corresponding to rearrangements of quarks inside hadrons) results in the $\psi(2S)K^+\pi^-$ final state. In this case, various "cusps" in the $\psi(2S)\pi^-$ invariant mass distribution may appear due to thresholds associated with the different excited charm and anticharm mesons. In order to examine the resonance nature of the $Z(4430)^-$ state, LHCb performed an additional analysis in which the Breit-Wigner description of the $Z(4430)^-$ lineshape is replaced by a function composed of an interpolation between six independent complex numbers at fixed $m(\psi(2S)\pi^-)^2$ positions. In this way, the lineshape itself can be determined from the data. The right-hand image in Figure 8 shows the results for the six complex numbers obtained from a fit, plotted in the Argand plane for values of $m(\psi(2S)\pi^-)^2$ increasing along the line from the topmost point. The red line superimposed on these points is the expectation for a Breit-Wigner amplitude which matches the data well, strongly supporting the hypothesis that the $Z(4430)^-$ is a resonant state.

It therefore seems clear that the $Z(4430)^-$ must be exotic with a minimal quark content of $c\bar{c}d\bar{u}$, but as before there is a major open question as to its internal structure. Observations of the same state in different decay modes would provide useful information. Belle have reported a significant contribution from the $Z(4430)^-$ in the $B^0 \to J/\psi K^+\pi^-$ decay, and it will be interesting to see if this is confirmed by LHCb. It is also important to establish if there are isospin partners, but this involves study of the $B^+ \to \psi(2S)K^+\pi^0$ decay, which has not yet been performed (final states involving $\pi^0$ mesons tend to be more challenging experimentally, in particular for LHCb). Another interesting approach is to explore the $B$ meson decays to a charm meson-anticharm meson pair and a kaon, where it should be possible to search for both charged and neutral states, albeit with reduced sample sizes.

As shown by the red and purple boxes in Figure 5, numerous other neutral and charged tetraquark-like states have been observed. A particularly interesting case is that of the charged $Z(3900)^\pm$ state that has been seen by BESIII, Belle and CLEO-c experiments as an enhancement in the $m(J/\psi\pi^\pm)$ spectrum of the $e^+e^- \to J/\psi\pi^+\pi^-$ process at centre-of-mass energy near 4.2 GeV (in the original BESIII and Belle analyses, the energy was tuned to be at the peak of a previously observed structure known as the $Y(4260)$, which itself is thought possibly to have an exotic origin). The mass of the $Z(3900)^\pm$ is approximately 24 MeV above the threshold for production of a $D^0 + D^{*-}$ meson pair. Subsequent study by BESIII of the $e^+e^- \to \pi^+ D^0 D^{*-}$ process has indicated a strong enhancement near threshold in the $m(D^0 D^{*-})$ spectrum, which is highly likely to be due to open-charm $Z(3900)^\pm$ decay. BESIII has also determined the quantum numbers to be $J^P = 1^+$, and reported observation of the neutral isospin partner $Z(3900)^0$ state in both $J/\psi \pi^0$ and $D^+ D^{*-}$ modes, with masses and widths in good agreement with those observed for the $Z(3900)^\pm$. Intriguingly, these states have yet to be observed in the decays of *b*-quark hadrons; in particular, there is no signal for the $Z(3900)^\pm$ in the Belle analysis of $B^0 \to J/\psi K^+ \pi^-$. Even with this substantial experimental information on the $Z(3900)^\pm$ states, there are conflicting theoretical models suggesting that they may be molecules of open-charm mesons, combinations of tightly-bound diquarks or some effect of the decay kinematics and closeness of the open-charm threshold.

---

³ The alert reader will note that the name of this particle is not very consistent with the latest measurements of its mass. The original name is still used, however, as the community is so familiar with it that a change would be likely to cause confusion.

As things stand, the intensively studied $X(3872)$, $Z(3900)^\pm$ and $Z(4430)^-$ states are the exception rather than the rule in this new era of exotic spectroscopy: the majority of claimed exotic states have only been observed by a single experiment in a single production and decay process. For example, the $Z_1(4050)^-$ and $Z_2(4250)^-$ states have only been seen by Belle as enhancements in the $\chi_{c1} \pi^-$ invariant mass in the decay $B^0 \to \chi_{c1} K^+ \pi^-$ (a BaBar analysis was not able to confirm the presence of exotic contributions). With large samples now available at LHCb and the Belle II experiment starting data taking in 2018 hopefully it will not be long before these more information on these and other states is available and, if they are confirmed, their masses, widths and quantum numbers measured from detailed amplitude analyses.

**Pentaquarks**

As discussed earlier, no convincing evidence of pentaquark states stood the test of time in over fifty years of exploration. However, in 2015 the LHCb experiment made a compelling claim of a heavy charmonium pentaquark that contributes to the $\Lambda_b^0 \to J/\psi p K^-$ decay. This process involves contributions from many conventional excited $\Lambda^*$ baryons (with quark content *uds*) decaying to $pK^-$, interference between which could potentially lead to spurious signals. Nonetheless, inspection of the invariant mass of the $J/\psi p$ system clearly shows signs of a narrow peak around 4.4 GeV (Figure 9 left), which could be due to a charged resonance with minimal quark content $uudc\bar{c}$.

To confirm this hypothesis, LHCb physicists constructed a multidimensional amplitude model that was fit to the data, building on experience from the study of the $Z(4430)$ state in the $B^0 \to \psi(2S) K^+ \pi^-$ process. In this case, the presence of spin-½ particles ($\Lambda_b^0$, $p$) in the initial and final states means that six dimensions are required to fully describe the kinematics of the process. Fourteen different $\Lambda^*$ baryons were included in the model, with parameters (masses, widths and $J^P$) set to their known values, but this was found to be insufficient to describe the data. Satisfactory fits were only achieved once two exotic $P_c^+ \to J/\psi p$ components were added: a lower mass state at ~4380 MeV with large width of

~205 MeV and a higher mass state at ~4450 MeV that is much narrower at ~39 MeV. Both states are needed, at high statistical significance, to describe the data. The right panel of Figure 9 shows the Argand diagram for the $P_c(4450)^+$ state, which clearly exhibits the expected resonant behaviour. The case is not so clear for the wider state and further studies will be required with larger data samples.

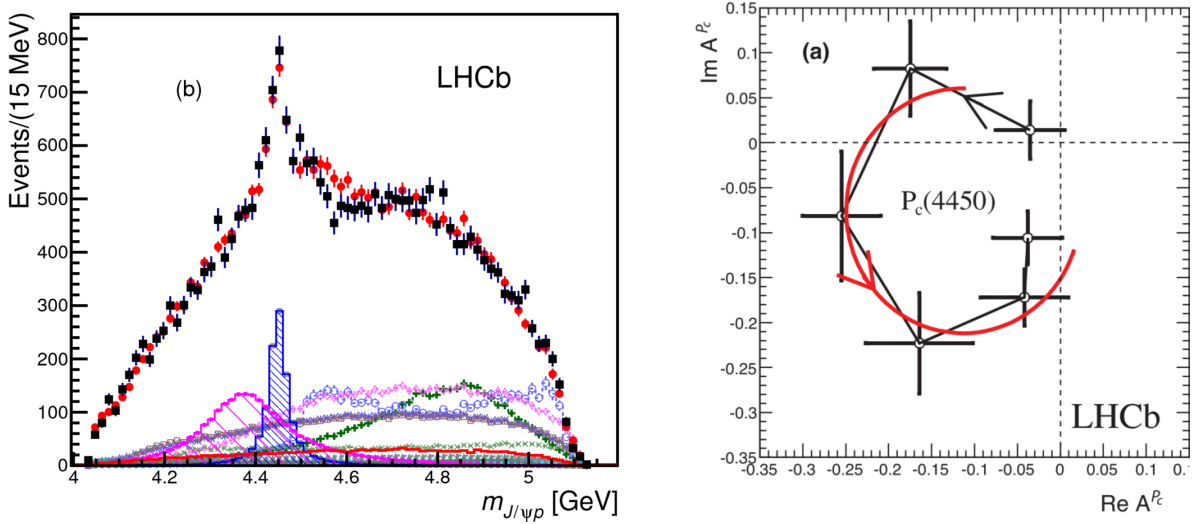

**Figure 9.** (left) Invariant mass spectrum of $J/\psi p$ combinations in $\Lambda_b^0 \to J/\psi p K^-$ decays. The hatched pink and blue histograms show the contributions from the $P_c(4380)^+$ and $P_c(4450)^+$ pentaquark states, respectively. The black points represent the data, which agree well with the total result of the fit shown in red. Remaining coloured points show contributions from different $\Lambda^*$ resonances. (right) The Argand diagram of the $P_c(4450)^+$ state.

Despite the highly significant signals, the quantum numbers of the pentaquarks could not be unambiguously determined from the amplitude analysis, with four possibilities giving acceptable fits to the data: solutions where the (lighter, heavier) states have spin of (3/2, 5/2) or (5/2, 3/2) with parity (−,+) or (+,−). Since knowledge of the quantum numbers is crucial to aid interpretation of the results, this needs to be addressed with further studies. One possibility is that the inclusion of more LHCb data and improved understanding of the $\Lambda^*$ baryon contributions will allow $J^P$ measurements of both pentaquarks. Another is that new production and decay mechanisms will provide insight. LHCb has started on this path by studying the $\Lambda_b^0 \to J/\psi p \pi^-$ decay mode. Since this mode as a smaller branching fraction than the discovery $\Lambda_b^0 \to J/\psi p K^-$ channel, the sample available for analysis is reduced by more than a factor of twenty. Nonetheless, it was possible to build a six-dimensional amplitude model to test the hypothesis that the data are consistent with only having contributions from excited nucleon resonances, $N^* \to p \pi^-$, or if exotic contributions (either $P_c^+ \to J/\psi p$ pentaquarks or $Z^- \to J/\psi \pi^-$ tetraquarks) are needed to describe the data. The conclusion from the LHCb study is that some combination of exotic hadrons appears to be needed, but at a level of significance that is not yet compelling. More data will be required to confirm which, if any, exotic states are contributing to the decay.

As mentioned for other exotic hadrons, it is always important to have confirmation of discoveries by independent experiments. Unfortunately, there are limited opportunities for experiments other than LHCb to contribute to the study of pentaquark states. Experiments based on the $e^+e^- \to Y(4S) \to B\bar{B}$ production process have a centre-of-mass energy too low to produce $\Lambda_b^0$ baryons. The high energy hadron colliders, the Tevatron and LHC, produce $b$ baryons copiously, but only the LHCb detector provides the capability to separate efficiently final states involving protons from the more copious background modes with pions. For this reason, the CDF, D0, ATLAS and CMS experiments have so far not been able to obtain results on pentaquark production. An exciting possibility is that it should be possible to produce pentaquarks in a completely different way, called photoproduction. The idea is that, since the photon has the same quantum numbers as the $J/\psi$, pentaquarks can be produced when a photon beam collides with protons in a target. Several experiments at the Jefferson Laboratory CEBAF facility are looking to exploit this opportunity. Since the experimental environment is rather clean, it may be possible to infer pentaquark production by detecting only the produced $J/\psi$, but the sensitivity will be enhanced if the proton can also be reconstructed. Although the photoproduction method has still to be demonstrated, if successful this approach will provide a novel avenue for pentaquark studies and could completely exclude the hypothesis that the $P_c^+$ states are caused by kinematic effects.

Assuming that the pentaquarks are true multiquark states, a familiar question again raises its head: how are the constituents bound? As with many exotic states, the presence of nearby two-hadron thresholds appears to play an important role in understanding their properties. For example, the $P_c(4450)^+$ state has a mass very close to the threshold for production of a $\chi_{c1}$ charmonium meson and a proton. This has spurred much speculation that the $P_c(4450)^+$ may arise due to the precise kinematics of the $\Lambda_b^0 \to J/\psi p K^-$ decay. One way to clarify this situation is to search for pentaquarks in the related $\Lambda_b^0 \to \chi_{c1} p K^-$ channel. This mode has recently been observed by LHCb, which bodes well for a future multidimensional amplitude analysis. Given that the $P_c^+$ states have quark content $uudc\bar{c}$ it is interesting to search for pentaquarks with open-strangeness ($udsc\bar{c}$) that could be observed as $J/\psi \Lambda$ resonances in $\Xi_b^- \to J/\psi \Lambda K^-$ decays. It is also important to explore possible decays to charm hadron pairs, such as $\Lambda_b^0 \to \Lambda_c^+ \bar{D}^0 K^-$. Although these modes are not as experimentally accessible as the golden channel for pentaquark discovery, $\Lambda_b^0 \to J/\psi p K^-$, the available sample sizes at LHCb are increasing rapidly, so there are good prospects for new results in the coming years.

**Doubly heavy baryons**

One of the difficulties of interpreting the data on exotic hadrons is the large uncertainties associated to theoretical predictions for their masses in different models. Thus, further studies of convention hadrons that can help to reduce these uncertainties will be important to enable progress. One particularly interesting area is that of doubly heavy baryons, *i.e.* baryons containing two heavy quarks. While modelling of baryons with three light quarks suffers from the three-body problem, familiar from classical gravity, doubly heavy systems are much simpler to describe as the two heavy quarks are almost static at the centre of the baryon, with the lighter quark orbiting around them. This facilitates predictions in many theoretical approaches, but perhaps most notably in diquark models. In this case, the constituents of a double charm baryon would be considered to be $[cc]u$, $[cc]d$ or $[cc]s$.

Until 2017 there was no clear observation of any doubly heavy baryon. The SELEX experiment at Fermilab had reported a signal of the $\Xi_{cc}^+$ ($ccd$) state, but with properties that were quite different from expectation. The state was not confirmed subsequently by any other experiment, and so it remained unclear whether it was truly a doubly heavy baryon. Since hadron collisions produce large quantities of all species of hadrons, there was a clear opportunity for experiments at the LHC to clarify this situation. However, the production of doubly heavy systems requires both heavy quarks to be produced kinematically close to each other, such that they can bind and hadronise, leading to large uncertainties on the production rates. Nonetheless, the observation of the $B_c^+$ ($c\bar{b}$) meson, containing the heavy $c$ and anti-$b$ quarks, at measurable production rates in both the Tevatron and LHC provided reasonable grounds for optimism that doubly heavy baryons could be observable.

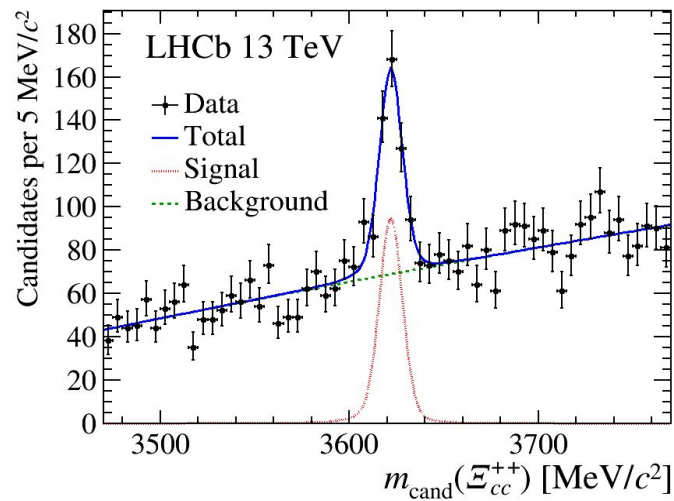

**Figure 10**. The signal for a double-charm baryon emerges from the invariant mass distribution of candidates obtained from data collected by the LHCb experiment. The result of a fit to the mass distribution is superimposed, showing how the red signal component is separated from the green background. Animation available online at http://iopscience.iop.org/book/978-0-7503-1593-7

Figure 10 shows the result of an analysis by LHCb searching for the doubly charged $\Xi_{cc}^{++}$ ($ccu$) state through its decay to $\Lambda_c^+ K^- \pi^+ \pi^+$. The animation (available online) shows the signal peak emerging as data are accumulated during 2016. Analysis of an independent data sample collected in 2012 confirms the signal. The mass is measured to be 3621.4 MeV with a relative precision of 0.02%. Although this is a different particle to the singly charged state claimed by SELEX, the results appear inconsistent since the masses differ by around 100 MeV, which is much larger than expected for isospin partners. The search for the counterpart $\Xi_{cc}^+$ state is therefore among several important follow-up measurements from LHCb, also including the lifetime and production rate of the $\Xi_{cc}^{++}$ baryon, that are hotly anticipated. It is also to be hoped, as ever, that results can be confirmed by at least one independent experiment – in this area, the prospects for the Belle II experiment appear promising.

21theoretical approaches, but perhaps most notably in diquark models. In this case, the constituents of a double charm baryon would be considered to be $[cc]u$, $[cc]d$ or $[cc]s$.

Until 2017 there was no clear observation of any doubly heavy baryon. The SELEX experiment at Fermilab had reported a signal of the $\Xi_{cc}^+$ ($ccd$) state, but with properties that were quite different from expectation. The state was not confirmed subsequently by any other experiment, and so it remained unclear whether it was truly a doubly heavy baryon. Since hadron collisions produce large quantities of all species of hadrons, there was a clear opportunity for experiments at the LHC to clarify this situation. However, the production of doubly heavy systems requires both heavy quarks to be produced kinematically close to each other, such that they can bind and hadronise, leading to large uncertainties on the production rates. Nonetheless, the observation of the $B_c^+$ ($c\bar{b}$) meson, containing the heavy $c$ and anti-$b$ quarks, at measurable production rates in both the Tevatron and LHC provided reasonable grounds for optimism that doubly heavy baryons could be observable.

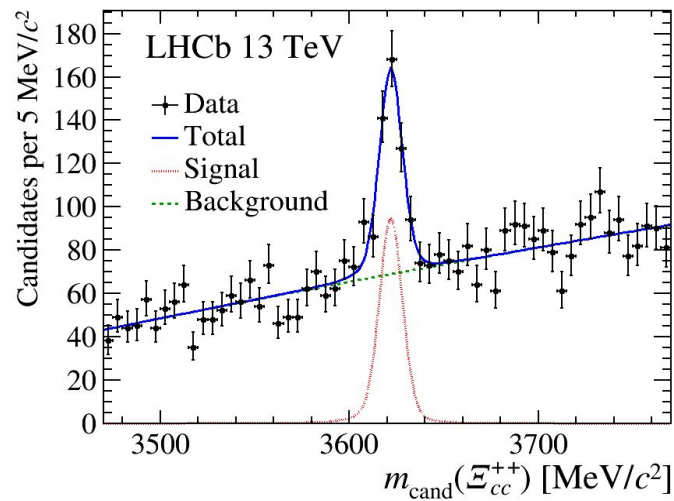

**Figure 10**. The signal for a double-charm baryon emerges from the invariant mass distribution of candidates obtained from data collected by the LHCb experiment. The result of a fit to the mass distribution is superimposed, showing how the red signal component is separated from the green background. Animation available online at http://iopscience.iop.org/book/978-0-7503-1593-7

Figure 10 shows the result of an analysis by LHCb searching for the doubly charged $\Xi_{cc}^{++}$ ($ccu$) state through its decay to $\Lambda_c^+ K^- \pi^+ \pi^+$. The animation (available online) shows the signal peak emerging as data are accumulated during 2016. Analysis of an independent data sample collected in 2012 confirms the signal. The mass is measured to be 3621.4 MeV with a relative precision of 0.02%. Although this is a different particle to the singly charged state claimed by SELEX, the results appear inconsistent since the masses differ by around 100 MeV, which is much larger than expected for isospin partners. The search for the counterpart $\Xi_{cc}^+$ state is therefore among several important follow-up measurements from LHCb, also including the lifetime and production rate of the $\Xi_{cc}^{++}$ baryon, that are hotly anticipated. It is also to be hoped, as ever, that results can be confirmed by at least one independent experiment – in this area, the prospects for the Belle II experiment appear promising.

However, in order to use doubly heavy baryons to reduce theoretical uncertainties due to QCD models, it will be necessary to obtain data on more than just one or two states. Observations of excited double-charm states are likely to require significantly larger data samples than are currently available, but may become within reach in the next five to ten years, both at LHCb and Belle II. Observations of doubly heavy baryons containing both beauty and charm quarks may also be possible on a similar timescale. Results on such states can be compared to those for $B$ mesons, including the $B_c^+$, and their excitations, to understand the effectiveness of the diquark model. It currently appears extremely challenging to observe a double-beauty baryon at any foreseeable experiment, but it would be unwise to discount the possibility of experimental innovations that could improve the sensitivity.

## 4. Outlook

**Models of exotic hadrons and how they can be distinguished**

Descriptions of the birth of the Standard Model often refer to a "particle zoo", reflecting a time at which many new states were observed but no organising principle had yet emerged to explain the spectroscopy. This was resolved in 1964 when Gell-Mann and Zweig independently proposed what became known as the quark model. This theory has stood the test of time remarkably well, and the discovery of exotic hadrons can be considered as a verification of one of its predictions. Nonetheless, it seems that we have a new zoo of exotic hadrons, and are again in a situation where a clear understanding of the internal structure of these particles is lacking.

Among the different models that have been proposed, the two that receive most attention are those involving molecules and compact multiquark states. Although we have referred throughout this document to both as examples of exotic hadrons, most people would agree that the former is less novel than the latter. The first hadron molecule, the deuteron, was discovered in the 1930s and is now used in a wide range of applications, most famously in nuclear reactors but also in areas such as medical imaging. It may be argued that if a proton and neutron can bind together through pion exchange, there is nothing particularly surprising in finding other hadrons doing likewise. Nonetheless, a strength of the molecule model is that it makes some simple predictions. Since the pion carries isospin, it can only bind hadrons which also have $I \neq 0$; moreover as a pseudoscalar ($J^P = 0^-$) it cannot bind two other pseudoscalars due to conservation of parity. All molecular states should have masses close to the threshold for the two bound hadrons, and corresponding quantum numbers. This model works reasonably well for some observed exotic hadrons, including the *Z(3900)* and similar states seen in the $b\bar{b}$ system, as well as the *X(3872)*. However, in each case there are certain caveats, *e.g.* the need for an admixture of conventional $c\bar{c}$ in the *X(3872)* wavefunction. Moreover, the molecular model cannot describe all the new states shown in Figure 5.

The compact multiquark picture, on the other hand, predicts a plethora of exotic states. The predicted masses tend to have fairly large uncertainties, but expectations for certain mass differences between related states are more precise. This model can successfully explain the pattern of masses between the *X(3872)*, *Z(3900)* and *Z(4430)* states, for example; it can also explain the common origin of the two observed pentaquark states. However, this model also predicts additional states that have not yet been observed. In some cases, such as the putative isospin partners of the *X(3872)*, the experimental limits present a significant challenge to the compact multiquark model.

If neither of these models, nor others that we do not discuss, can alone describe all data, is it possible that the exotic states are explained by some combination of the different types? Certainly there is no rule in QCD to forbid that both molecules and compact multiquarks exist, so what are the counter-arguments? One is that it might appear unusual for these different types of exotic particles to start appearing in experiments at around the same time. However, this can be explained by the dramatic improvements in sensitivity that were achieved by the BaBar and Belle experiments in the early 2000s, and by other experiments listed in Table 2 since then. Another is that it seems contrary to Occam's razor: the idea that the simplest explanation for a puzzle is often the correct one. This, however, is not a rigorous argument. As things stand, a composition of the different types of possible conventional and exotic hadrons, all emergent from the underlying theory of QCD, seems the most plausible explanation of the data, but further progress in both experiment and theory will clarify matters.

There are two particularly exciting developments in theory that may bear fruit in the near future. The first is that the observation of the $\Xi_{cc}^{++}$ baryon allows the diquark model to be refined and to make reasonably precise predictions for $[bb][\overline{ud}]$, $[bc][\overline{ud}]$, and $[cc][\overline{ud}]$ tetraquarks. There have even been suggestions that experiments would ultimately be able to find $[cc][\overline{cc}]$, $[bb][\overline{cc}]$ and $[bb][\overline{bb}]$ states, the last being a truly "beauty-full" tetraquark. Observations of such states would be an impressive validation of the model. The second is that more lattice QCD calculations including all relevant operators are expected to become available. A first attempt has been made to investigate the structure of the *X(3872)* in this way, with results that suggest it has components of both $c\bar{c}$ and $D\overline{D}^*$, but not of a compact tetraquark. It will be extremely interesting to see if improved lattice QCD calculations support this conclusion, and if further predictions can be made.

**Experimental prospects**

Since the discovery of charmonium, a reasonable approximation has been that the available data samples increase by an order of magnitude every 5 years. While this clearly cannot continue indefinitely, there are good prospects for at least the next decade. The LHC and its experiments will continue to run at CERN, with upgrades giving access to much larger data samples than currently available. The BEPC collider is likewise continuing to operate, collecting data at relevant energies for studies of $c\bar{c}$ states. The Super-KEKB accelerator is undergoing commissioning at the time of writing, and is expected to start to deliver large quantities of data to the Belle II experiment around the end of 2018. The experimental programme at JLab is also ramping up, with potential to provide novel insights into pentaquarks as well as other areas of QCD. Meanwhile, a new facility, FAIR, is being constructed at GSI in Germany. There, a beam of antiprotons will be used to study exotic hadrons in the PANDA experiment. In contrast to high-energy hadron colliders, PANDA will have precise control of the energy of the antiproton beam which should enable a detailed study of the *X(3872)* lineshape, for example. PANDA will also be able to study high-spin states that are less accessible at other facilities.

Many of the most important measurement to be performed have been mentioned in earlier sections. These include studies of new production and decay modes of established exotic hadrons, and searches for their isospin partners. It is also expected that new exotic hadrons may be discovered by broadening the range of final states that are investigated. One important area is to study decays with open charm hadrons in the final state. Another is to supplement studies of modes containing *J/ψ* or *ψ(2S)* mesons with other charmonia. While the majority of *b*-hadron decays to a *J/ψ* and two oppositely charged tracks (pions, kaons or protons) have been studied, there are rather few results to date involving related channels with $\eta_c$ or $\chi_c$ states. Being more experimentally challenging, these require the larger data samples that will become available.

All the established exotic hadrons so far contain either a $c\bar{c}$ or $b\bar{b}$ core. It is not clear whether this is an essential feature of exotic hadrons, or whether, for example, tetraquarks with four different quark flavours could exist. Although D0 has published a claimed observation of an *X(5568)* state decaying to $B_s^0 \pi^\pm$, which must therefore contain *b*, *s*, *u* and *d* quarks or antiquarks, this has not been confirmed by other experiments with apparently better sensitivity. Nonetheless, continued searches for this and similar types of exotic hadrons are well motivated.

The field of exotic hadron spectroscopy has until now been driven by experimental results, and it is likely that this will continue. It is therefore important that current and future experiments work together on a coherent and coordinated programme of research to pin down the properties of all states in the exotic spectrum and, importantly, determine the links between them. Together with continued progress in theory, this will allow us to understand the true dynamics underlying the structure of exotic hadrons. Achieving this goal will lead to improved knowledge of QCD, and help to address a central enigma of the Standard Model.

## 5. Additional resources

- For very accessible and pedagogical introductions to particle physics we recommend *Introduction to Elementary Particles* by D. Griffiths (Wiley, 2008) and *Modern Particle Physics* by M. Thomson (Cambridge University Press, 2013). These texts cover all of the basics of the Standard Model of particle physics, both from a theoretical and experimental point of view.
- For those interested in the origins of the quark model of hadrons we recommend reading the 1964 papers from M. Gell-Mann (*Phys. Lett.* **8**, 214 (1964)) and G. Zweig (reprinted in *Developments in the Quark Theory of Hadrons*, Hadronic Press, 1980), where they describe the concept of combining quarks (or *aces* in the case of Zweig) together to form composite mesons and baryons. An open access version of Gell-Mann's paper is available from http://tuvalu.santafe.edu/~mgm/Site/Publications_files/MGM%2047.pdf.

- Readers who are keen to have a crack at solving QCD and claiming the Millenium prize can find more information about it at http://www.claymath.org/millennium-problems/yang--mills-and-mass-gap.
- The field of exotic hadron spectroscopy is rapidly developing and the wealth of new experimental observations and theoretical interpretations is vast. To keep track of these advancements there have been several recent review articles that summarise the research and have extensive bibliographies to the original scientific papers. We particularly like those from M. Karliner and colleagues (to appear in *Ann. Rev. Nucl. Part. Sci.*), S. Olsen and colleagues (*Rev. Mod. Phys.* **90** (2018) 015003), A. Ali and colleagues (*Prog. Part. Nucl. Phys.* **97** (2017) 123) and R. Lebed and colleagues (*Prog. Part. Nucl. Phys.* **93** (2017) 143). Open access versions of these reviews can be downloaded from https://arxiv.org/abs/1711.10626, https://arxiv.org/abs/1708.04012, https://arxiv.org/abs/1706.00610 and https://arxiv.org/abs/1610.04528, respectively.
- If you would like to discover more about the controversy over various experimental claims for pentaquarks in the light-quark sector of QCD then we highly recommend the excellent review article *"On the conundrum of the pentaquark"* by K. Hicks (*Eur. Phys. J.* **H37** (2012) 1). An earlier, non-restricted, version of this article can be downloaded here https://arxiv.org/abs/hep-ph/0703004. A short review of non-$q\bar{q}$ mesons, focussing on possible glueball and hybrid candidates in the light-quark sector, forms part of the 2017 Review of Particle Physics http://pdg.lbl.gov/2017/reviews/rpp2017-rev-non-qqbar-mesons.pdf.
- The images used in Figure 3 showing two possibilities for the internal dynamics of a pentaquark (bound combination of meson and baryon or five tightly bound quarks) were produced by CERN. The original press release can be found here https://press.cern/press-releases/2015/07/cerns-lhcb-experiment-reports-observation-exotic-pentaquark-particles.
- The original version of Figure 5 came from S. Olsen (*Front. Phys.* **10** (2015) 121), which also acts as a shorter review article of hadron spectroscopy. An open access version is available from https://arxiv.org/abs/1411.7738.
- The original paper that claimed discovery of the *X(3872)* tetraquark (and the source for Figure 7) is by the Belle collaboration (*Phys. Rev. Lett.* **91** (2003) 262001). This is the most highly-cited paper from the Belle collaboration, which is itself remarkable given that the experiment was designed primarily to study *CP*-symmetry violation in the *B* meson system. The quantum numbers of the *X(3872)* have subsequently been determined with high statistical significance by the LHCb collaboration (*Phys. Rev.* **D92** (2015) 011102). Open access versions of these results can be downloaded from https://arxiv.org/abs/hep-ex/0309032 and http://arxiv.org/abs/1504.06339, respectively.
- The first observation of the charged charmonium-like $Z(4430)^-$ state was made by the Belle collaboration (*Phys. Rev. Lett.* **100** (2008) 142001). A more complete analysis of the full information available in the $B^0 \to \psi(2S) K^+ \pi^-$ decay was performed by the LHCb collaboration (*Phys. Rev. Lett.* **112** (2014) 222002), which demonstrated for the first time the resonant-like behaviour of the $Z(4430)^-$ (see Figure 8). Open access

- versions of these papers can be downloaded from http://arxiv.org/abs/0708.1790 and http://arxiv.org/abs/1404.1903, respectively.
- The original paper on the discovery of the charmonium-like pentaquark states (and the source for Figure 9) is by the LHCb collaboration (*Phys. Rev. Lett.* **115** (2015) 072001). This is one of the most highly-cited of over 400 publications from LHCb. All results from LHC experiments are published under gold open access arrangements, and therefore this paper can be freely downloaded from the journal at https://doi.org/10.1103/PhysRevLett.115.072001
- The paper reporting the observation of the double-charm $\Xi_{cc}^{++}$ baryon is from the LHCb collaboration (*Phys. Rev. Lett.* **119** (2017) 112001; https://doi.org/10.1103/PhysRevLett.119.112001). The original source for the animation of the $\Xi_{cc}^{++}$ invariant mass (Figure 10) is http://lhcb-public.web.cern.ch/lhcb-public/Welcome.html#Xicc.
- An interesting recent example of science in action is given by the publications from the D0 collaboration that claim first evidence for the *X(5568)* state, which would be a tetraquark containing four different flavours of quark. Similar studies have been performed by the LHCb, ATLAS and CMS collaborations at the LHC and D0's competitor experiment at the Tevatron, CDF. None of them find any sign of the *X(5568)* state in their data. It remains to be seen how this story will evolve with time and more data being collected. The open-access versions (and published journal reference, where available) of these results can be found at http://arxiv.org/abs/1602.07588 (*Phys. Rev. Lett.* **117** (2016) 022003), http://arxiv.org/abs/1608.00435 (*Phys. Rev. Lett.* **117** (2016) 152003), http://arxiv.org/abs/1712.06144, http://arxiv.org/abs/1712.09620, http://arxiv.org/abs/1712.10176 and http://arxiv.org/abs/1802.01840 .